# Separation and Electrical Properties of Self-Organized Graphene/Graphite Layers

Manuel R. Mailian[1], Aram R. Mailian[2, a]

[1] *LTX-Credence Armenia, 2 Adonts str., 0014, Yerevan, Armenia.*
[2] *Institute for Informatics, 1 P. Sevak str., 0014, Yerevan, Armenia.*

[a] Corresponding author: amailian@ipia.sci.am

**Abstract.** Intrinsic layered structure of graphite is the source of ongoing and expanding search of ways of obtaining low-cost and promising graphite thin layers. We report on a novel method of obtaining and seperating rubbed graphite sheets by using water soluble NaCl substrate. The electrical behavior of sheets was characterized by current–voltage measurements. An in-plane electrical anisotropy depending on rubbing direction is discovered. Optical microscopy observations combined with discovered non-linear electrical behavior revealed that friction leads to the formation of sheet makeup which contain an optically transparent lamina of self-organized few-layer graphene.

## INTRODUCTION

Since the Scotch tape separation of two-dimensional single-layer graphite crystals [1] different approaches of straightforward separation were developed for obtaining single- and few-layer graphene. The easiest way, gettimg thin graphite structure by rubbing, has received relatively little attention. The reason is probably that these layers or "paper electronics" were disbelieved to offer predictable and reliable physical characteristics for using in actual electronics.

On the other hand, because of expected physically interesting characteristics there is still a growing interest in new simple fabrication methods of graphite nanoscale thick layers with predictable physical properties and amenable to external perturbations. Recently, a number of researchers reported that these layers expose interesting properties and behave like a solid-state material with predictable semiconducting properties. Their piezoresistive effect [2], higher mechanical flexibility [3] [4], stability [5], and other attractive properties [6] enable the use of rubbed off layers in sensors. From basic study point of view, peculiarities of layer formation, temperature dependent carrier transport and Raman spectra suggested modified electron gas in such layers [7]. In particular, it was suggested that in layers modified by rubbing a lamina of few-layer graphene is formed due to self-organization.

In this study we report on innovative and versatile technique of seperating graphene/graphite structures obtained by rubbing. By optical microscopy we characterized the structure and stacking order of these layers and in conjunction with I-V measurement revealed their peculiar electrical behavior.

## EXPERIMENT

We obtained graphite sheets by repeatedly rubbing parent graphite bulk on NaCl substrate. Then we immersed the sandwiched sheet/substrate structure into water and after dissolution of NaCl substrate detached graphite sheet floated freely in water and reached water surface (Figure 1). The sheet was transferred on a glass substrate in either of two positions: back-side-down (BSD) or back-side-up (BSU) (shown in bottom callout in Figure 1(a)).

Phenomenologically defined, during rubbing interface layers of both parent bulk and rubbed off on the substrate are slid, cleaved, transferred and pressed by a combined mechanical action which is effectively to be described as *Combined Mechanical Modification (CMM)*.

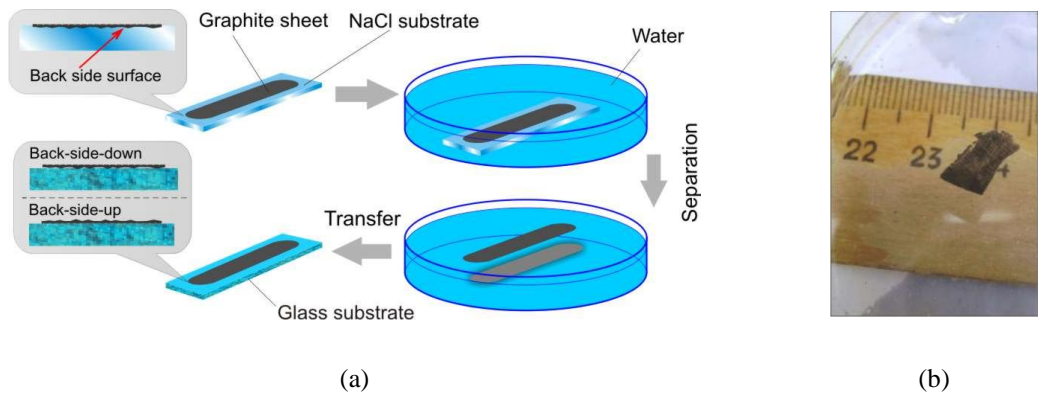

(a)          (b)

**FIGURE 1.** (a) Schematic presentation of separation of CMM layers. The sheet was transferred in either BSD or BSU position. (b) A sample of separated CMM sheet floating in water.

For optical observation an AmScope B490B microscope was used. The optical images were taken in two modes: transmission and reflection. Non-polarized white light was used for illumination. The angle of side illumination in reflection mode was taken close to 20° to the structure plane. Electrical measurements were carried out using two-probe method using contacts gently pressed on the structure surface.

## RESULTS AND DISCUSSION

CMM sheets exhibited a particularly peculiar electrical behavior reflecting the phase of evolution of surface morphology. The electrical resistance of a CMM sheet measured along rubbing path strongly depends on the number of rubbing, N (Figure 2(a)). The initial conducting layers of the sheet (N=3-4) for a typical sample as presented in Figure 2(a) was ~ $10^3$ times less conductive than the sheet shaped after multiple rubbing (N>20).

In addition, the electrical behavior diverges depending on the rubbing direction (Figure 2(b)). I-V characteristics obtained along rubbing path is close to linear while across rubbing path is obvious non-linear with large slope at zero bias (Figure 2(b)).

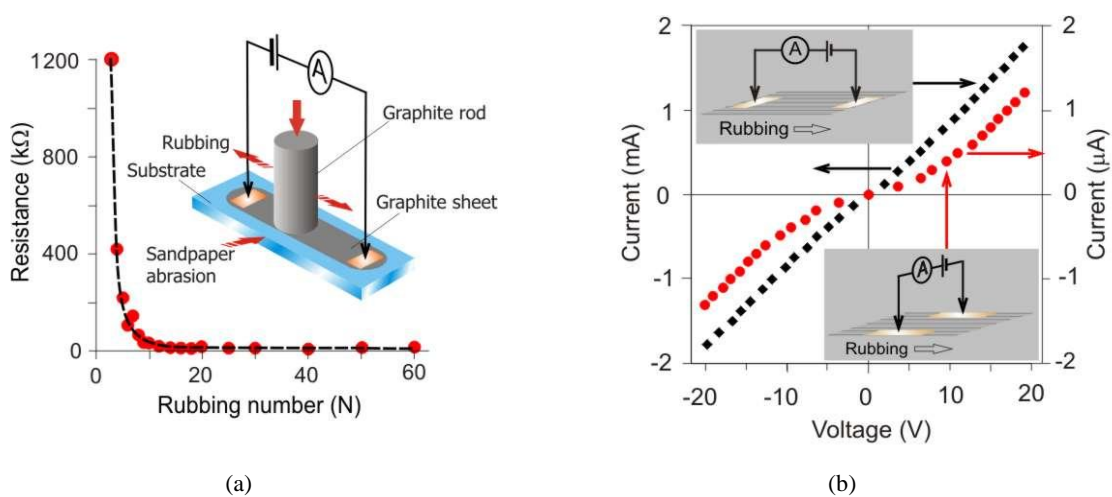

(a)          (b)

**FIGURE 2.** (a) The CMM sheet resistance vs. Rubbing number measured along rubbing path. Inset shows the process of rubbing on NaCl substrate. (b) Typical I-V characteristics of CMM layers along (black diamonds) and across (red circles) rubbing path. Insets show the schematics of measurement.

Optical microscopy revealed detailed pattern of surface morphology of both sides of a CMM sheet and its specific through-thickness stacking (Figure 3). When processing of the substrate surface and rubbing of graphite rod was performed in a straight line, the transmission images of CMM sheets contained two sets of parallel dark strips as shown in optical microscopy images (Figure 3(b), (d)): 1) along rubbing path ($R_f$) and 2) along sandpaper dug trenches on NaCl substrate surface ($R_s$). Characterizing topography of transmission images in terms of thickness, it is obvious that dark strips are the planar projections of thicker structures in a CMM sheet.

On the other hand vast surface area remained bright even after manifold rubbing as shown in Figure 3(b). We observed that these areas are darker than bright area of naked substrate. As should be expected, the topography of transmission images remained the same after transfer (Figure 3(d)). Usually friction-shaped dark strips are thinner. Interestingly, in case of paper substrate the dark traces imitated worm-like celluloid threads of paper tissue.

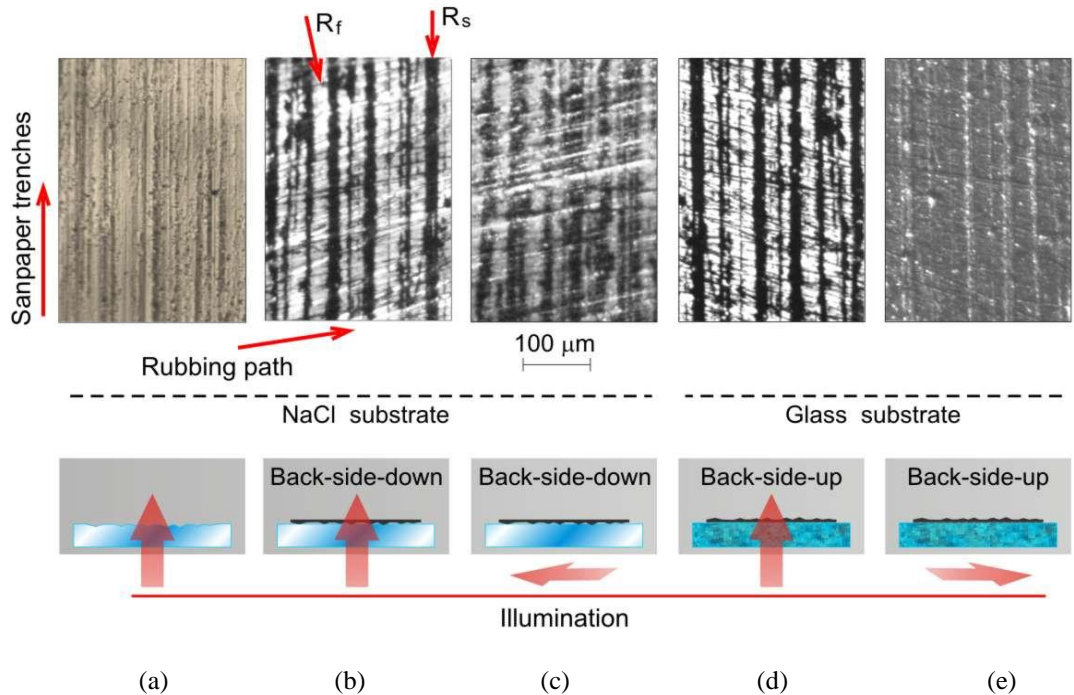

**FIGURE 3.** Optical microscopy images of a fragment of a CMM sheet with rubbing number N>20. (a) NaCl surface processed by sandpaper. (b) The fragment of transmission image of the layer rubbed on NaCl substrate. The sheet is in intrinsic BSD position. (c) Reflection image of the same fragment under side illumination. (d) Transmission image of the same fragment of transferred sheet. The image is the mirror reflection of the image (b) since the sheet was transferred in BSU position. (e) Reflection image of transferred sheet. Red arrows in bottom panel show the direction of illumination.

The morphology of opposite surfaces of a CMM sheet is basically different and as expected should reflect formatting mechanisms. The back-side or bottom surface is formed due to filling graphite flakes the unevenness on the substrate surface under pressure. Its surface morphology is the inverse reproduction of substrate surface topography (upper callout in Figure 1(a)). The morphology of the top surface is developed mainly due to friction forces emerging between the bulk graphite and earlier rubbed off CMM layers and hence should bear the traces of rubbing. Consequently the morphology of the sheet bottom is substrate-born while the top surface is friction-modified.

The conductance measurements show that the morphology of the sheet surface directly affect the electrical properties of a CMM sheet. First, the electrical conductivity along any dark strip regardless the origin - substrate-born or friction-modified - is ~$10^3$ times higher and exhibits Ohmic I-V behavior. This is well interpreted by the assumption that dark strips are projections of structures thick enough to behave as a bulk conductor. Meanwhile, I-V characteristics measured across dark strips is non-linear with larger slope at zero bias (Figure 2(b)). This type of behavior is characteristic of and peculiar to the graphene.

Further observations show, that the majority of dark strips seen in transmission images (Figure 3(b), (d)) turn into shining streaks in reflection mode (Figure 3(c), (e)). We distinguished two sets of such streaks. First one strictly matched the projections of friction-modified structures when as-grown sheet on the substrate was examined (intrinsic BSD arrangement) (Figure 3(c)). The other set was observed on the surface of BSU transferred sheet and matched substrate trenches (Figure 3(e)). Since bright streaks are seen due to the reflection from surface roughness the above transformation clearly indicates that we observed ridge-and-valley structures on both sides of the sheet. The height of substrate-born ridges was estimated to be about the size of sandpaper grains (≤50μm) and friction shaped ones ≤5μm.

To further study, we cut off a piece of a sheet projected of alternating dark strips and measured I-V characteristics through transparent area (Figure 4(a)). The sheet was positioned BSU allowing mounting electrical contacts with the substrate-born thicker conducting ridges (marked as Rs in transmission image). As can be seen the electrical behavior across dark strips is obvious non-Ohmic (Figure 4(b)).

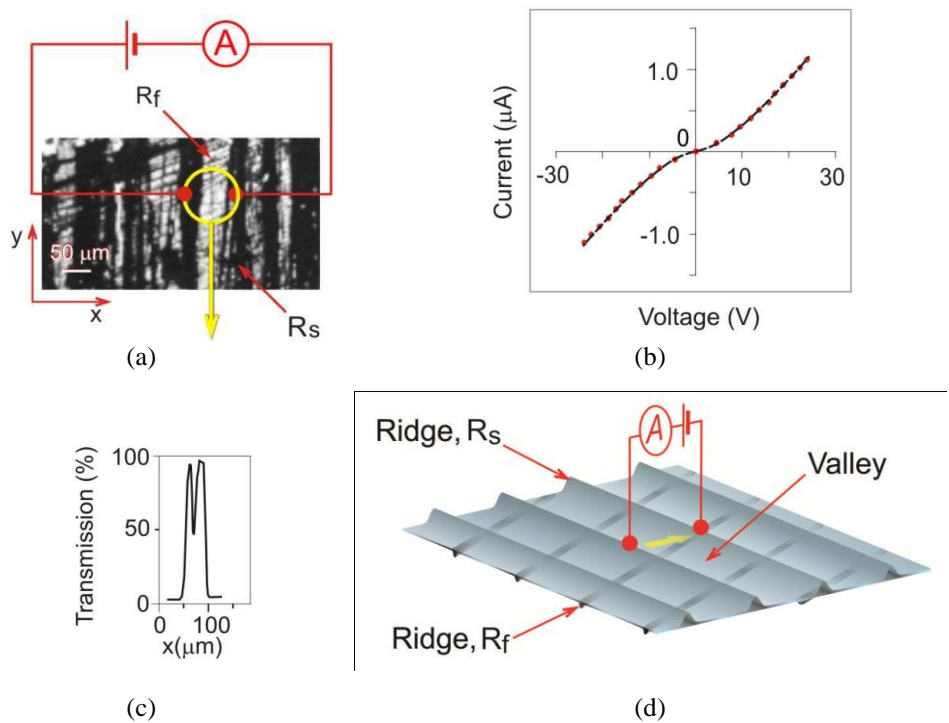

(a) (b)

(c) (d)

**FIGURE 4.** Measurement of electrical characteristics of randomly chosen transferred segment of a CMM sheet. The image is taken in transmission mode. The sheet was transferred upset, i. e. rubbed surface looks opposite to the reader. (a) Experimental setup for measuring conductance. $D_s$ indicates the lines originating from trenches on NaCl substrate and $D_r$ stands for dark lines on the structure originating from rubbing. (b) Optical transmission across x line in the area included in yellow circle. (c) I-V characteristics obtained between dark lines. (d) Model of CMM sheet in BSU arrangement.

In transmission projection, dark strips at an angle toward x-y axes ($R_f$) connect the dark strips ($R_s$) through the bright area. Since the sheet segment projected as transparent area between conducting ridges shows non-Ohmic behavior it means that ridges ($R_f$ and $R_s$) are weakly coupled electrically. That means that both sets of high conducting ridges on either sides of the CMM sheet are isolated by less conductive sandwiching lamina. To estimate the thickness of transparent area we calculated its optical transmission, which reached 95-97 % in brighter parts of transparent area. (Figure 4(c)). Furthermore, since its electrical behavior is not Ohmic we suggest that transparent areas are fragments of a wider spread few-layer graphene (Figure 4(d)).

Reflection images strongly support this assumption. Under side illumination only friction-shaped dark strips turn into shining streaks (Figure 3(c)), while in BSU position only substrate-born dark strips turn into shining streaks (Figure 3(e)).

Proposed packing model explains the sharp decrease in resistance along rubbing path measured between the contacts attached to sheet surface (Figure 2(a)). On rubbed face the density of friction-modified parallel dark strips increases with rubbing hence the total resistance of parallel connected conducting ridges (Rf) should sharply drop rubbing.

As for the process of rubbing, it is well-known that an atomically thick layer shaped during rubbing is a natural product of any tribological system [8]. To possible explanation of formation of CMM structures and their physical behavior note that the only physical influence in rubbing is realized through forces of friction. The forces of friction evolve the crystal layers in friction zone into a atomically thin self-organized protective crystal structure with highly ordered lattice. As for carbon, by means of molecular dynamics simulations it is shown that in diamond, the hardest allotropy of carbon, two-layer graphene lamina is likely to emerge due to sliding under pressure [9]. As a result, the layers with quantum-size thickness ought to essentially change the behavior of electron gas in such 2D crystalline lamina which should exhibit non-Ohmic characteristics.

## CONCUSIONS

In conclusion, the approach of using water soluble NaCl substrate was successfully used for separating graphite layers obtained by rubbing. The optical microscopy observation in correlation with non-linear electrical characteristics revealed that layered sheets consist of substrate-born and friction-modified high conductive ridge-like structures emerging on both sides of the of a flat lamina of high-resistive few-layer graphene.

## REFERENCES

1. K. S. Novoselov, A. K. Geim, S. V. Morozov, D. Jiang, Y. Zhang, S. V. Dubonos, I. V. Grigorieva and A. A. Firsov, Science **306**, 666-669 (2004).
2. T-L. Ren, H. Tian, Xie D, Y. Yang, Sensors **12**, 6685-94 (2012).
3. N. Kurra, D. Dutta, G.U. Kulkarni, Physical Chemistry Chemical Physics **15**, 8367-72 (2013).
4. Y. Wang, H. Zhou, Energy & Environmental Science. **4**, 1704-7 (2011).
5. P. Mandal, R. Dey, S. Chakraborty, Lab on a Chip **12**, 4026-8 (2012).
6. K. ul Hasan, O. Nur, M. Willander, Appl. Phys. Letters **100**, 1104-1106 (2012).
7. A. R. Mailian, G. S. Shmavonyan, M. R. Mailian ArXiv e-prints p. 3929 (2014).
8. I. S. Gershman and N. A. Bushein, "Elements of Thermodynamics and Self-Organization during Friction," in *Advanced Surface-Engineered Materials and Systems Design*, edited by G. S. Fox-Rabinovich & G. E. Totten (CRC Press, 2006), pp. 13-56.
9. M.M. Van Wijk, A. Fasolino, ArXiv e-prints p. 1404.3492 v1 (2014).